# Observation of spin Nernst photocurrents in topological insulators

[1]T. Schumann, [1]N. Meyer, [2]G. Mussler, [2]J. Kampmeier, [2]D. Grützmacher, [3]E. Schmoranzerova, [4]L. Braun, [4]T. Kampfrath, [1]J. Walowski, [1]M. Münzenberg

1. Institut für Physik, Universität Greifswald, Felix-Hausdorff-Straße 6, 17489 Greifswald, Germany
2. Peter Grünberg Institut (PGI-9) and Jülich-Aachen Research Alliance (JARA-FIT), Forschungszentrum Jülich, 52425 Jülich, Germany
3. Department of Chemical Physics and Optics, Charles University, Ke Karlovu 3, 121 16 Prague, Czech Republic
4. Department of Physical Chemistry, Fritz Haber Institute, Faradayweg 4-6, 14195 Berlin, Germany

**Abstract: The theoretical prediction of topological insulators in 2007 triggered tremendous interest. They are of fundamental interest because of their topological twist in k-space, which comes along with unidirectional, spin-polarized surface-state currents, required for spin-optoelectronics. This property makes topological insulators on one hand perfect materials for optically generated, ultrafast spin-bunches spin-current sources for the generation of THz radiation. On the other hand, those spin-polarized surface-state currents when generated by a voltage lead to large spin Hall effects, or when generated by a temperature gradient to the thermal analogue, the spin Nernst effect. Both mutually convert charge/ heat currents into transverse spin currents leading to spin accumulations**. **By connecting both research fields, we show the evidence of heat-transport related spin Hall effects that can be extracted from opto-transport experiments. This heat-driven spin Nernst effect drives a transverse spin-current and affects the optical spin-orientation in the three-dimensional topological insulator. This manifests as a modification of the circular polarization-dependent photocurrent. We illuminate the detailed thermocurrent distribution, including the influence of edges and contacts, in spatially resolved current maps.**

## Introduction

A three-dimensional topological insulator (3D-TI) is a material with strong spin-orbit interaction that gives rise to surface states with specific properties [1, 2, 3, 4, 5]. One of the most prominent is that, in contrast to the insulating bulk of the material, the surface states display a metallic-like conductivity [6]. Furthermore, the surface states show a distinct spin polarization that connects the direction of the carrier spin to its momentum (spin-momentum locking). Recently, large effort has been aimed, at using the spin-momentum locking for generation of charge currents with a large degree of spin polarization, pure spin currents, or to control a charge current direction in the surface channel [7]. In the seminal work of the Gedik group [4], this effect was applied to excite a helicity dependent photocurrent at the surface at room temperature [4, 8, 9]. Different effects of thermal origin can influence the transport in these materials, affecting the helicity-dependent signals. Therefore, it is crucial to understand the origin of each contribution to the photocurrent and their underlying microscopic mechanisms. The relative complexity of the photocurrent behavior arises from the broad energy regime of the optical excitation used in the seminal studies. The prototype 3D-TI materials are chalcogenides, such as $Bi_2Se_3$, $Bi_2Te_3$ or $Sb_2Te_3$ [1, 3]. From the intensive materials exploration and studies of photoemission over the last years, it is possible to gain control over their band structure and the gap crossing Dirac cone. All of them are narrow-gap semiconductors and have bulk band gaps in the range of 200 – 300 meV. Their Fermi level position can be adjusted over a wide range, in alloys such as $(Bi, Sb)_2Te_3$, resulting in very low carrier concentrations and low intrinsic conduction of the bulk material [10]. Since the band gap is very narrow with respect to optical applications, the excited states are inside the Dirac cone, only for excitation wavelengths larger than 4 µm. However, typical photon energies used in these optical experiments are in the visible and near infrared (NIR) region, at around 0.8 µm, and the photon energies are five to eight times the value of the bulk band gap. Therefore, the surface states in the Dirac cone are excited far into the bulk valence band. This gives rise to a whole variety of optically induced effects. Depending on the time scales relevant for the effect under study, two types of experiments are performed, (i) the ultrafast photocurrent generated in the dynamic regime by femtosecond laser pulses, which allows separating the different time scales of charge and spin relaxation, and (ii) quasi-equilibrium carrier transport by continuous wave excitation, typically performed in a form of an opto-electrical measurement on micro patterned devices [7].

The ultrafast spin and carrier dynamics has been successfully investigated in many types of time-resolved optical experiments, by direct detection of helicity dependent polarization in films using magneto-optics [11], by THz emission [12] or by electric sampling using amorphous Silicon-Auston switches [9, 13]. In our previous work [11], we used the all-optical pump-probe technique to examine the coherent ultrafast spin-dynamics of $(Bi,Sb)_2Te_3$ (BST) alloys. The electrons are excited into the second Dirac cone [14] and consequently relax towards the bottom of the conduction band [15]. The spin polarization of the electrons at the moment of excitation is given by the energy and polarization of the absorbed photon. One finds spin-dependent photogalvanic effects arising from spin-orbit interaction in non-centrosymmetric materials [16], and the photon drag effect [17] or [18] arising from a band-dependent speed of the charge carriers. These effects consequently result in spin-polarized currents in a semiconductor after optical excitation. All the above-mentioned phenomena can be of importance with respect to the quasi-equilibrium photocurrents detected in transport experiments, and respectively have impact on the photocurrents. Very recently, in the work by Pan et al. [19], these phenomena were disentangled by studying photocurrents over a large wavelength range, combined with a modification of the electrochemical potential by a gate voltage. Apparently, the photocurrents can have also a complex spatial dependence. Substrate effects can contribute, as well as the depth dependent photo-Dember effects, which is a typical source of ultrafast spin currents in semiconductors [5]. At the contacts, a spatial inhomogeneity of the induced photocurrents is expected. The origin of this inhomogeneity can be manifold, for example the modification of the local potential landscape at metal/semiconductor interface and the formation of a Schottky barrier. This has recently been demonstrated for ultrafast photocurrents in graphene nanojunctions [20, 13]. Therefore, in our work, we want to discuss the lateral effects by looking in detail into their spatial dependence along the Hall bar.

Apart from that, this establishes another way to create spin-polarization by sending a current though a thin film in a per se nonmagnetic material. The spin Hall effect converts a longitudinal charge current in the bulk of the material into a transverse spin current. The efficiency is given by the so called spin Hall angle $\theta_{Spin\ Hall} = \frac{|J_{spin}|}{|J_{charge}|}$ [7]. The spin Hall effect is exploited in heavy metals [7, 21] to create large spin orbit torque at interfaces in ferromagnet bilayers [22, 23]. A search for increasing the efficiency is ongoing. So far, the spin Hall angle $\theta_{Spin\ Hall}$ is far below one in these materials, typically about 0.1 – 0.3. For ferromagnet/topological insulator heterostructures in contrast, very high spin orbit torque efficiencies have been reported recently [24, 25, 26, 27, 28]. The benchmark of topological insulators is the large intrinsic spin-orbit interaction, resulting in much higher charge-to-spin conversion than in heavy metals. Huge factors of $\theta_{Spin\ Hall}$ have been observed in structures with magnetically doped TIs [24], and even values of >18 for sputtered granular topological insulator films [25]. One explanation is that those large values result from the definition. A large resistance of the bulk of TIs compared to metals, makes the denominator in $\theta_{Spin\ Hall}$ close to zero, increasing the value of $\theta_{Spin\ Hall}$. This shows the relevance of a detailed understanding of the mechanisms leading to the extreme spin orbit torque switching efficiency [24, 25, 26]. The spin Hall effect has a thermal analogue, the spin Nernst effect. A longitudinal heat current in the material drives a transverse spin current. Microscopically it originates from an interplay of temperature driven electrons, feeling locally spin-orbit effects in the material. This effect has been only recently discovered in Pt [29] but not yet in topological insulators. We want to understand how light generated effects, instead of voltage generated currents, interact with surface currents in TIs. Light acts in two ways on the TI film, it creates (i) a photo-excitation of the carriers and (ii) a heat gradient. The question is, can we drive a spin Hall effect by light to accumulate spins? In this work we aim to combine the spin Hall effect physics in TIs, with the optical excitation of the materials in terms of a spin Hall photoconductance [30]. We report on the appearance of a spin Nernst effect-like signal in the photocurrent in a Hall-bar. We attribute this signal to a spin accumulation originating from a spin-orbit driven transverse spin-current, as a response to the longitudinal thermocurrent, which is detected via a helicity-dependent photocurrent modified by the spin accumulation.

**Results**

*Experimental Setup*
In the following we describe the performed experiments. The laser radiation with a wavelength of 785 nm is focused down to a 3 µm Gaussian beam on the sample surface, yielding a power density of 0.6 W/cm² (linear signal range). To increase signal separations and achieve relaxed signal levels, the laser output is modulated by a rectangular function at a frequency in the 100 Hz range. In this slow regime the photo-response mirrors the quasi static rectangular excitation pattern, and at the same time excludes dynamic effects, which occur at higher modulation frequencies. The setup allows precise laser spot positioning by stepper motors, with high axial and lateral resolutions down to 200 nm along the sample surface, while measuring the reflectivity and the voltages generated by the photocurrents. We have developed a method to map the contributions to the generated voltage for each point spatially and as a function of the polarization of light. For this purpose, at each point, the reflectivity and the photovoltage $V_{\text{photo}}$ generated by the thermocurrent, excited by the laser beam, is recorded for a full rotation of the quarter-wave plate (qwp). We use the same analysis, as proposed by Ganichev in reference [17] for the analysis of spin photogalvanic effects in GaAs, and which was also used by McIver et al. in reference [4] for the study of topological insulators. We fit each point by the following equation:

$$V_{\text{photo}}(\theta) = C \cdot \sin(2\theta) + L_1 \cdot \sin(4\theta) + L_2 \cdot \cos(4\theta) + D$$

The angle $\theta$ indexes the iterative turning of the qwp around the light propagation axis in a perpendicular plane for each single measurement, as depicted in Fig. 1. When rotating the qwp, the light polarization changes from linear polarized photons for 0°, 90°, 180° and 270° to right-circular for 45°, 225° and left-circular for 135°, 315° polarized light. That means, the equation reflects the voltage contributions changing with the angle $\theta$ 's periodicity as a function of polarization variation: the voltage measured at the contacts as the angle $\theta$ of the qwp is rotated is divided into four contributions to extract the parameters D, $L_1$, $L_2$ and C. The factor D gives the polarization independent voltage. This arises from the Seebeck voltage generated by heat gradients driven by the laser. The components $L_1$ and $L_2$ have the period of $4\Theta$ and they have a fixed phase of 90° relative to each other. They depend on the angular alignment of the incident light, its polarization relative to the samples surface in general, and especially to crystalline axes of the material. So both potentially address multiple effects connected to the linear polarization axis, which consist of common effects like polarization sensitive absorption, changing effectively the heat gradient, and less common effects like a linear dependent photon drag effect [18]. The factor C that has the period of $2\Theta$, represents the contribution of interest here: it originates from a circularly polarization-driven photocurrent, e.g. the difference in photocurrent for right- and left-circular polarized light helicity. The circularly polarized light promotes the excitation of one spin species spin up or down respectively. It generates a directional current which originates from the imbalance in presence of spin-momentum locking in the topological insulator. This drives a directional current on the surface. The process is called circular photogalvanic effect. To avoid the experiments being affected by dichroic absorption effects that would also modify the factor C amplitude, we measure the reflectivity and photocurrents at the same time for a full rotation cycle. These two-dimensional power absorption and reflectivity maps allow determining the influence of local heating by the scanning laser spot. Since the absorption is polarization dependent itself, this is a precondition to interpret the background of Seebeck voltage that contributes in addition to the polarization dependence. Fig. 1a shows the experiment, a sketch of the setup with an optical microscope image of the sample and the sample geometry. The laser beam angle of incidence is kept fixed at 45° to the sample normal. In Fig. 1b, a reflectivity map of the sample area is depicted, taken simultaneously with the voltage map. The Au contacts show the highest reflection in white, the substrate appears almost black on the scale (low reflectivity signal) and the topological insulator is grey in between. These reflectivity and $V_{photo}$ maps are also compared for s- and p- polarization in detail in the section below. Thereby, we thoroughly analyze possible edge absorption effects. We find them to be minor for these local measurements.

In the following discussion we have chosen a $(Bi_{0.57}Sb_{0.43})_2Te_3$ ternary heterostructure (BST) with a total thickness of 16 nm. BSTs allow the tuning of the Fermi level between valence and conduction band, based on the bismuth to antimony ratio, especially as close as possible to the Dirac point [10]. In addition, this doping in the band gap minimizes the conducting charge carrier concentration of the thin film. The minimum charge carrier concentration of $3 \cdot 10^{13}$ cm$^{-2}$ is determined by transport measurements. The ternary $(Bi_{0.57}Sb_{0.43})_2Te_3$ alloy is grown by molecular-beam epitaxy [31, 32] on a Si-(111) surface, which is the upper most layer of silicon-on-insulator (SOI) substrate. The BSTs are analyzed by various experiments, including x-ray photoelectron spectroscopy (XPS), x-ray diffraction (XRD), Raman spectroscopy, angular resolved photo electron spectroscopy (ARPES). ARPES monitors the Fermi level movement through the band gap by regulating the antimony amount, which is also confirmed by the corresponding Hall effect mapping of the n- to p-type doping transition. Our samples are chosen to be located very close to this compensation point with the lowest intrinsic conductance. After structuring the thin films into Hall-bars by optical lithography and Ar$^+$-ion dry etching they are stored under dry vacuum conditions. This surface reconstruction procedure is verified by ARPES measurements to yield the correct surface states and Fermi level again.

After electrical wire bonding, large-scale overview maps are performed on the Hall bar devices to analyze the role of the contacts and substrate on the photocurrents. The data is presented in Fig. 2. Each point of these maps has been analyzed by equation (1) to extract the parameters D, $L_1$, $L_2$ and C. D the thermovoltage, and C the photogalvanic voltage, are displayed in Fig. 2a and b, respectively. The contacts on the right side are used during the experiment to pick up the signals in between. It can be seen that the largest polarization independent voltages appear, when the laser spot excites the media right at the structured contacts, but also finite and non-trivial dependencies along the topological insulator structure are found. The thermovoltage D signals reverse the sign at each contact and the signal strength is of the size of a few mV at the contacts, elsewhere mainly they are in the dimension of hundreds µV. These thermovoltage distributions can be explained by the lateral variations of the reflectance, shown in Fig. 1b. The largest values of D are found in particular at the topological insulator to substrate interface close to the contacts. Almost no thermovoltage is found for the excitation on top of the Au metal contact, where the highest reflectance occurs. These voltage signals show a sign reversal for the doping of the topological insulator going from n- to p-type which is expected. The center of our study is the search for the part of the signal sensitive to the exciting photon's helicity, which is associated with the parameter C in equation (1) the circular photogalvanic voltage associated with the helicity dependent currents. These signals are at least a factor of hundred smaller in amplitude compared to parameter D, associated to the thermocurrent (Fig. 2b). This makes their extraction generally laborious. In the large overview map, we see that with moving the excitation laterally abrupt changes from +5 to -5 µV for C can appear, for substrate to TI edges close to the contacts. We tested s- and p-polarized light, photocurrent maps in overview maps and in detail in the Hall bar center for significant changes at edges or other potential sensitive regions (e.g. plasmonic effects). However, these changes turn out to be negligible and will not be discussed in the following. The spatial irregular distribution of the values for C tells us that no meaningful values can be extracted in the large overview scans for the dichroic shift currents.

In order to extract meaningful values for each parameter, photocurrent scans with higher spatial resolution are required. Fig. 3 shows measurements, focused on the geometrical center of the Hall bar device between two contacts, using a higher spatial resolution. Again, we discuss the extracted values for C and D. Fits are exemplarily shown for three positions 1-3 in Fig. 3b. This position is chosen to reduce the thermocurrent. Its minimum is expected in the geometrical center between the contacts. Consequently, the non-thermal contribution to signals driven by the circular polarization driven photocurrent will be dominating. This idea is similar to the concept used by McIver et al. [4] previously. The 2D map in Fig. 3 in the upper left shows the thermocurrent distribution for the chosen region. For all points, the average thermovoltage is positive and in the order of 125 µV, indicated by the red color scale over the full scan range. Again, for a more detailed discussion, averaged profiles consisting of multiple lines of a section of the data area, are shown parallel and perpendicular to the Hall bar, to the top and right side of the color plot (Fig. 3): The marked regions are two perpendicular lateral profiles, where we identify a monotone drop of the D value along the Hall bar. This is expected when moving the laser spot between contacts. However, although there is a linear dependence, due to a thermovoltage offset on the BST Hall bar, the data shows no zero crossing. Since the values of $L_1$ and $L_2$ go with sine and cosine of $4\Theta$, the complex L-value can be plotted as modulus and its phase. For the modulus we find values about a factor of ten smaller than D. The phase change is the strongest for s- and p- polarization, as shown in Fig. 5. For C values extracted from the fits at each point of the scan regions, we find that the circular photo galvanic effect is again a factor of hundred smaller, in the µV range. The most important finding from this section is that there is a unidirectional offset thermovoltage present on the BST Hall bar, which indicates a well defined temperature gradient $\nabla T$ directed along the Hall bar device.

The unexpected findings are presented in Fig. 4, where a spatially resolved map of the C-value, originating from the difference in photo-voltages for the right-and left circularly polarized laser excitation, is shown in a false color plot. In the average region, the C-values stay almost constant along the bar, which is contrary to the thermovoltage. In addition, a complex behavior occurs at the edges of the Hall bar. There, the voltage reverses direction as indicated by the blue and red areas at both edges. This resembles immediately a spin-Hall like signal in the Hall bar structure. In the profile line, these are more visible as local maxima and minima. We find an asymmetry of our photocurrent with respect to the polarization of the light for right or left circular polarization. The spins generated by the selective excitation by the circularly polarized light generate a positive or negative voltage depending on their polarization.

What is the origin of this signal that appears as a spin polarization accumulation at the edges, measured by the excitation difference for the two polarizations? Recently, related findings of a spin-Hall photoconductance effect were reported by P. Seifert et al. [30]. They observed that at the edges of $Bi_2Te_2Se$ platelets, a spin-accumulation modified by the photoconductance occurs. By varying the driving current, they tested different symmetry scenarios and deduced that the accumulation of spins at the edges arises from the spin-Hall effect in the TI bulk states. Further, they found that the spin accumulation at the edges gives rise to an optically induced kind of magnetoresistance effect, e.g. change of the edge channel photoconductance that is connected to the spin-population and the light's circular polarization. Here, we observe a similar signature, but without any application of an external current. The microscopic process is depicted in Fig. 4 schematically: instead of a voltage driving the transverse spin current, we observe a thermal gradient driving a transverse spin accumulation at the edges via the spin Nernst effect, the thermal analogue is the spin Hall effect [29]. This thermally driven spin polarization changes the weighting in the spin population, sensed by the exciting photon helicity. Thus, the laser lights action is twofold: the laser generates the heat gradient and the spins accumulated at the device edges, determined by the size of the spin Nernst effect. Additionally. the size of the photogalvanic effect itself depends now on the helicity depend-photoexcitation of the carriers. The latter appears as a contrast in the spin-voltage maps of the C value that originates from the difference for left and right circularly polarized light, sensing the accumulated spins at the device edges. The spin-density accumulated at the edges thus can be accessed. It originates naturally from the longitudinal temperature gradient, built in as an offset heat current along the Hall bar that drives a pure transverse spin current.

## Discussion

We have shown a study of the photo-excitation in the three-dimensional topological insulator $(Bi_{0.57}Sb_{0.43})_2Te_3$ with a fixed intrinsic doping level leading to the lowest intrinsic conduction of the bulk states. For the thin film geometry, the Hall bar device is studied in an overview map of the polarization dependent photocurrents to elucidate the effects especially of the contacts, revealing an offset thermovoltage along the Hall bar. The high-resolution map shows a contrast of the device at the edges that have the signature of the spin Nernst effect. We interpret this signal as a transverse spin current driven by the thermal gradient along the Hall bar. The resulting spin accumulation influences then the spin dependent circular photogalvanic effect that allows detecting a difference in the spin accumulation via the polarization dependent photovoltage signal. This gives new possibilities to use the spin-polarization created by thermal gradients and by light-induced effects to study the processes of spin-accumulation in topological insulators with respect to spintronic applications, magnetoresistance effects and spin torques in topological materials.

## Acknowledgements

We acknowledge funding through the Deutsche Forschungsgemeinschaft (DFG) via the priority program, SPP "Topological Insulators: Materials - Fundamental Properties - Devices" (SPP 1666) and DAAD-PPP Czech Republic, project "FemtomagTopo". We acknowledge the fruitful discussions with A. Holleitner.

## Author contributions

M.M. and E.S. designed the experiment. T.S., N.M., J.W., E.S., developed the interfaces to measure and analyze the experimental data and carried out the photocurrent experiments. M.M., T.S., J.W. and E.S. developed the model. G.M., J.K., and D.G. prepared and characterized the samples. T.S., N.M., G.M., J.K., D.G., L.B., T.K., J.W. and M.M. discussed the data and interpretation. T.S., E.S., G.M., J.W. and M.M wrote the manuscript. All the authors reviewed and discussed the manuscript.

Figures:

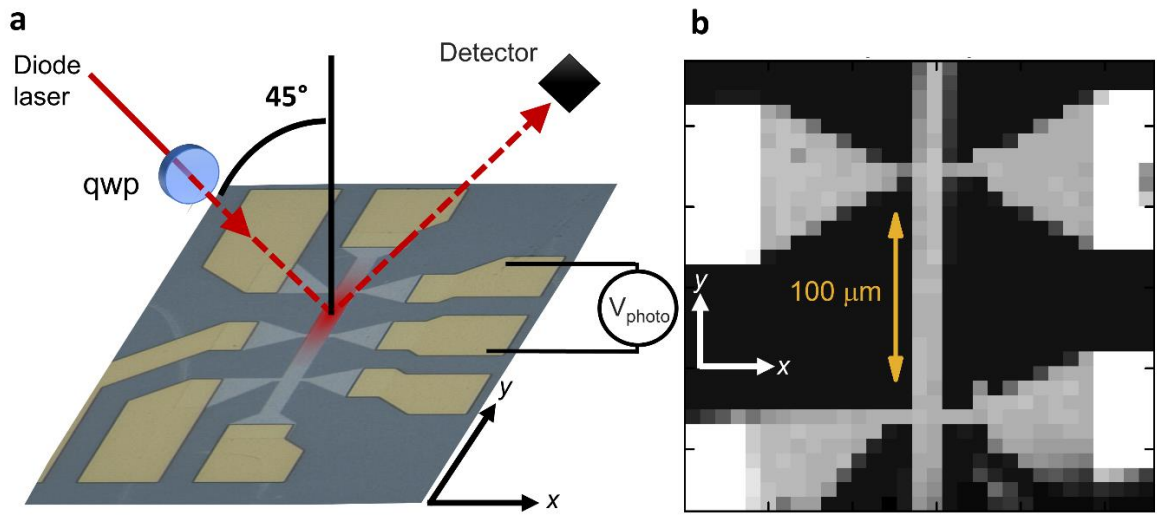

**Fig. 1| Experimental setup. a**, Schematic layout of the Topological insulator sample structure with intrinsic doping (16 nm $(Bi_{0.57}Sb_{0.43})_2Te_3$) and measurement alignment. The Au contact pads are the golden leads in the microscopy image. The topological insulator Hall bar structure is seen in lighter gray on the darker grey silicon substrate. The incidence angle of the incoming laser beam is 45° to the layer normal, shown by the solid and dashed red lines. Additionally, the heat gradient area is indicated schematically by the red area between two contacts. **b**, A reflectivity map derived from scanning the sample in x and y direction as indicated by the coordinate system. The investigated 100 µm long Hall bar region is clearly distinguishable from the contact pads and the substrate.

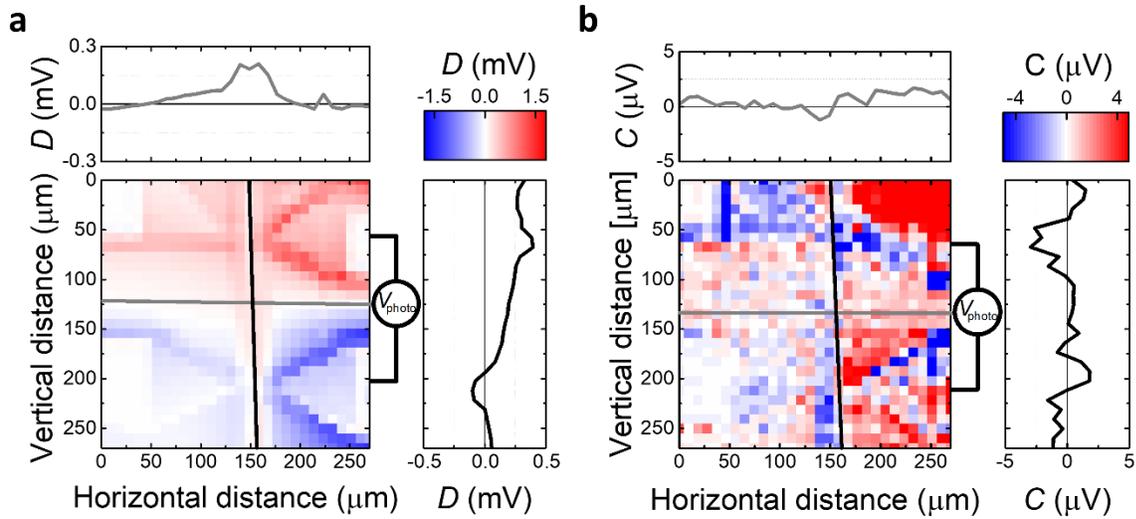

**Fig. 2| Overview, extracted parameters mapped to the position on the sample.** Both panels show the parameters $D$ and $C$ extracted by fitting equation 1, with s-polarized light modulated for a full quarter wave plate (qwp) rotation at each position, while scanning the laser spot over the sample surface. **a**, Thermovoltage D contribution is plotted in a false color plot. The profiles along the Hall bar (black line) show that D is always positive. **b**, The parameter C, represents the circular photogalvanic voltage contribution generated by the circularly polarized laser excitation.

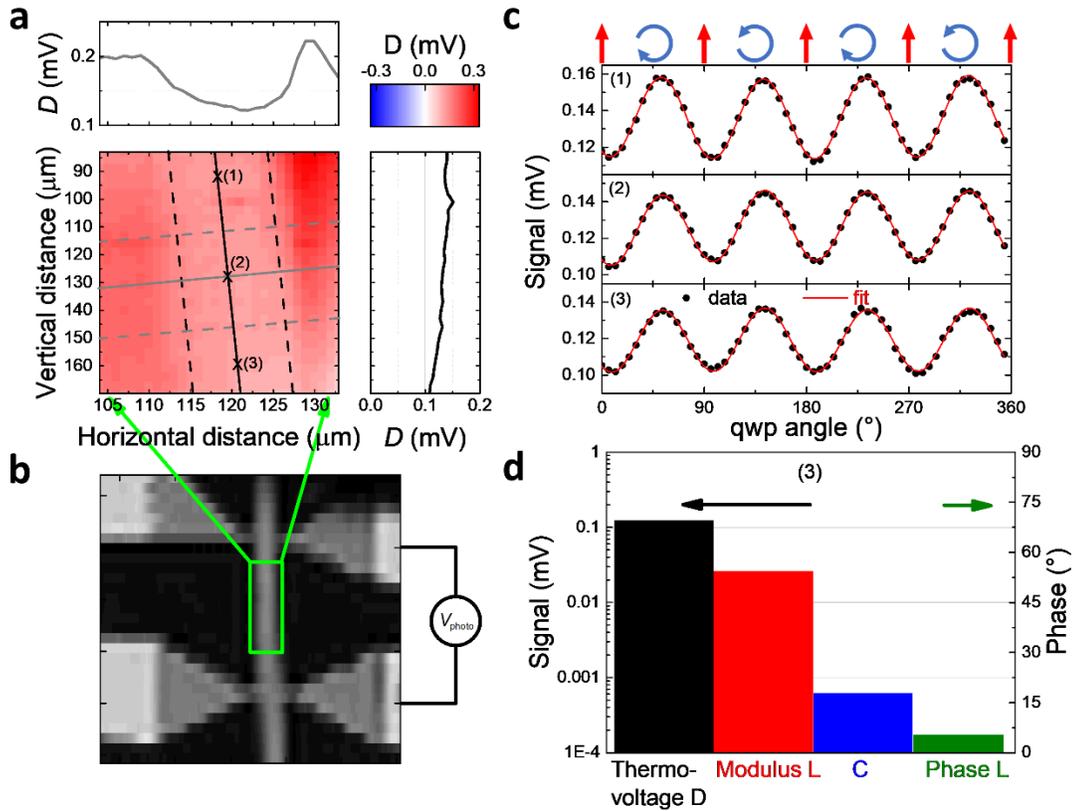

**Fig. 3| Detailed measurements and photogalvanic effect analysis. a**, A map showing the thermovoltage parameter D extracted from measurements with higher resolution within the area indicated in panel **b**. For the specific points (1), (2) and (3) the full polarization dependent signal (black dots in **c**) in respect with the quarter wave plate (qwp) rotation angle $\theta$ are plotted together with the curves obtained from the fits to equation 1. **d**, The parameters D, $L_1$, $L_2$ and C (for L modulus and phase related to real and imaginary part of the linear photogalvanic effects $L_1$ or $L_2$) for point (3) are shown in the bar diagram on the left axis scale, with the phase of L on the right scale.

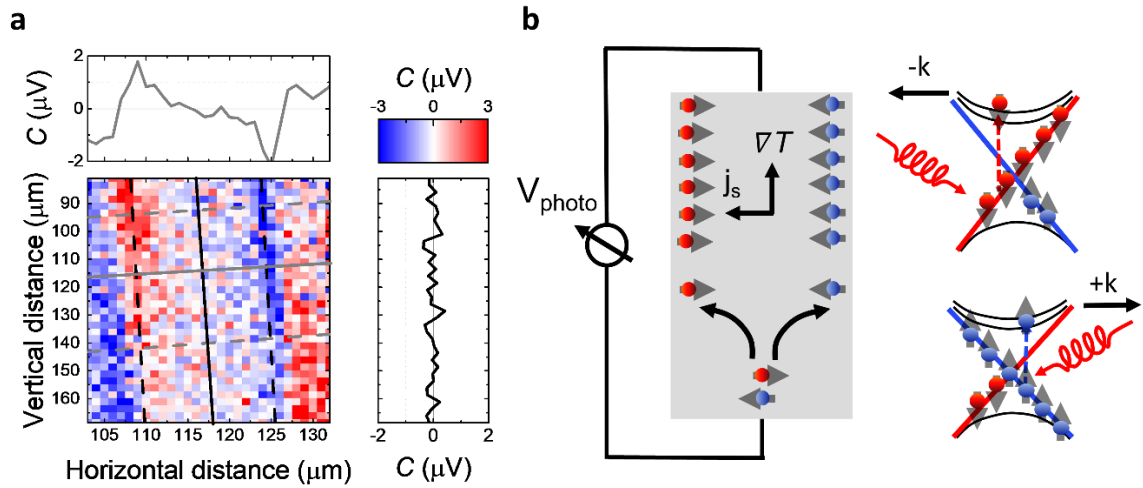

**Fig. 4| Interplay between the offset-thermocurrent and circular photogalvanic current signal.** Comparison of two optical excitation scenarios. The incident laser beam is **a** p-polarized before quarter wave plate (qwp) modulation. The panel show parameter C, the voltage generated by the circular photogalvanic effect, created by the circular polarized light for both scenarios. **b**, The model explains the twostep spin-accumulation process. First, the spin Nernst effect accumulates the spins towards the edges, perpendicular to the temperature gradient $\nabla T$. Second, the altered spin occupation modifies the circular photogalvanic current.